\documentclass[%
 reprint,
 amsmath,amssymb,
 aps,
]{revtex4-2}

\usepackage{graphicx}
\usepackage{dcolumn}
\usepackage{ragged2e}
\usepackage{bm}

\begin{document}

\preprint{APS/123-QED}

\title{Enhanced bandwidth in radiation sensors operating at the fundamental temperature fluctuation noise limit}

\author{Chang Zhang\textsuperscript{1}}

\author{Zachary Louis-Seize\textsuperscript{1}}

\author{Maxime Brazeau\textsuperscript{1}}

\author{Timothy Hodges\textsuperscript{1}}

\author{Mathis Turgeon-Roy\textsuperscript{1}}

\author{Raphael St-Gelais\textsuperscript{1,2}}\email{raphael.stgelais@uottawa.ca}
\affiliation{\textsuperscript{1}University of Ottawa, Department of Mechanical Engineering}
\affiliation{\textsuperscript{2}University of Ottawa, Department of Physics}

\date{\today}

\begin{abstract}
Temperature-based radiation detectors are an essential tool for long optical wavelengths detection even if they often suffer from important bandwidth limitations. Their responsivity, and hence their noise equivalent power (NEP), typically degrade at frequencies exceeding the cutoff set by their characteristic thermal response time ($\tau_\text{th}$), i.e., at $\omega > \tau_\text{th}^{-1}$. Here we show that this bandwidth limitation can be broken when a radiation sensor operates at its fundamental temperature fluctuation noise limit. The key enabler of this demonstration is a nanomechanical sensor in which frequency stability is limited by fundamental temperature fluctuations over an unprecedentedly large bandwidth of 54 $\text{Hz}$. In this range, the sensor performance remains within a factor 3 from its peak detectivity ($D_T^* = 7.4 \times 10^9~\mathrm{cm \cdot Hz^{1/2} W^{-1}}$) even though the thermal cutoff frequency is 30 times lower (i.e., $1/2\mathrm{\pi} \tau_\text{th} = 1.8~\text{Hz}$). We also derive and validate experimentally closed-form expression predicting maximum bandwidth enhancement in the context of nanomechanical resonators interfaced with a closed-loop frequency tracking scheme.

\end{abstract}

\maketitle


\section{Intro}
Temperature-sensitive nanomechanical resonators (NMRs)~\cite{zhang2013nanomechanical,laurent201812,blaikie2019fast,hui2016plasmonic,piller2022thermal,zhang2024highdetectivityterahertzradiation,zhang2016room,vicarelli2022micromechanical} have attracted significant interest as an alternative to traditional thermal sensing approaches that rely on electrical sensing (e.g., resistive bolometers~\cite{wang2012vanadium,renoux2011sub}, pyroelectric detectors~\cite{blaikie2019fast,hui2016plasmonic,piller2022thermal}, and thermopiles~\cite{hsu2015graphene,st2011single}). In principle, mechanical-based temperature detection in NMRs can provide immunity to electrical noise (e.g., Johnson–Nyquist noise) that limits performance in traditional thermal-based detectors. This immunity creates a path to reaching the fundamental detectivity limit of thermal photon fluctuation at room temperature $D_{T,photon}^{*} \approx 10^{10}~\text{cm} \cdot \text{Hz}^{1/2} \text{W}^{-1}$~\cite{ref1}.

Reaching this fundamental detectivity limit with NMR-based radiation sensors requires eliminating all non-fundamental sources of frequency fluctuations. In recent years, these non-fundamental fluctuations have been shown to originate in large part from thermomechanical resonator fluctuations, and from noise in the readout used to capture resonator vibration. Comprehensive studies of these noise sources can be found in~\cite{demir2019fundamental,demir2021understanding,sadeghi2020frequency,bevsic2023schemes}, which also outlines the central importance of including the frequency tracking schemes—such as phase-locked loop (PLL) or self-sustaining oscillators—when predicting frequency noise.

Despite these recent progress in noise modeling, demonstration of NMRs in which these noise sources are minimized below fundamental temperature fluctuation remain scarce. Previous work \cite{zhang2023demonstration} explored the modeling of thermal fluctuation noise in a PLL frequency tracking scheme with preliminary experimental validation within a limited frequency range. Other recent comparative analysis of silicon nitride nanomechanical resonators have approached the temperature fluctuation noise limit~\cite{kanellopulos2025comparative} but without reaching it clearly over a broad frequency bandwidth.

Here, we demonstrate NMRs with a frequency stability dominated by temperature fluctuation noise over an unprecedentedly large bandwidth of 54~Hz. We also provide updated modeling for temperature fluctuation noise within closed-loop frequency tracking scheme, thus greatly improving predictability of noise figure in this regime.

Building on this result, we unveil an important benefit of reaching the fundamental temperature fluctuation noise limit over such a large bandwidth. In this regime, we show that detectivity and noise equivalent power remain undegraded at frequencies greatly exceeding the resonator thermal responsivity cutoff (i.e., at $\omega > 1/\tau_{th}$, where $\tau_{th}$ is the sensor characteristic thermal response time).
\section{Theory}
When using NMRs to perform thermal radiation sensing, noise equivalent power (NEP) in $\mathrm{W \cdot Hz^{-1/2}}$ can be defined as:
\begin{equation}
\mathrm{NEP} = \frac{\sqrt{S_y}}{R}, 
\label{art2_eq1}
\end{equation}
where $S_y$ quantifies unwanted resonator frequency fluctuations, and $R$ quantifies the sensor responsivity, i.e., its sensitivity to absorbed radiation. We use unitless fractional frequency ($y = \delta \omega / \omega_r$) throughout this work such that $S_y$ is a noise spectral density (one-sided) expressed in units of $\mathrm{Hz^{-1}}$. Similarly, $R$ is in units of $\mathrm{W^{-1}}$ and is given by
\begin{equation}
R(\omega) = \frac{\gamma \alpha}{G} |H_{th}(\omega)|,
\label{art2_eq2}
\end{equation}
where $G$ is the total thermal conductance between the NMR and its environment, in $\mathrm{W/K}$, $\gamma$ is the optical absorption coefficient at target detection wavelength, and $\alpha$ represents the temperature coefficient of fractional frequency shift, in $\mathrm{K^{-1}}$. This coefficient $\alpha$ is mostly dictated by the NMR’s material properties and is discussed below (see Eq.~\ref{art2_eq17}) for the specific resonator used in this work. $H_{th}(\omega)$ accounts for NMR thermal response time $\tau_{th}$ under uniform illumination, and is defined as:
\begin{equation}
H_{th}(\omega) = \frac{1}{1 + j \omega \tau_{th}}.
\label{art2_eq3}
\end{equation}

For the general case of an arbitrary white noise $S_y$, we see from Eq.~\ref{art2_eq1} that NEP degrades rapidly at frequencies exceeding the resonator thermal response time (i.e., for $\omega > \tau_{th}^{-1}$) because of its inverse dependence on the thermal responsivity filter (Eq.~~\ref{art2_eq3}). However, this is not always true if we examine more closely the different possible contributions to $S_y$.

Here, we consider the three most common noise sources for the overall noise profile of NMRs, namely thermomechanical noise $S_{y,{tmech}}$, instrument readout noise $S_{y,{read}}$, and fundamental temperature fluctuation noise $S_{y,T}$:
\begin{equation}
S_y = S_{y,{tmech}} + S_{y,{read}} + S_{y,T}.
\label{art2_eq4}
\end{equation}
Among these three noise sources, $S_T$ is the universal limiting noise for thermal-based radiation sensors of all types~\cite{ref1} (e.g., bolometer, thermopile, pyroelectric detector), including NMRs. For NMRs, $S_{y,T}$ can be expressed as:
\begin{equation}
S_{y,T}(\omega) = \frac{4k_B T^2 \alpha^2}{G} \left| H_{{th,eff}}(\omega) \right|^2,
\label{art2_eq5}
\end{equation}
where $k_B$ is the Boltzmann constant and $T$ is the surrounding temperature. $H_{{th,eff}}(\omega)$ denotes the effective thermal response of the resonator, averaging all local thermal energy fluctuations over the sensor volume. As shown in~\cite{kanellopulos2025comparative}, the value of $H_{{th,eff}}(\omega)$ is close to $H_{th}(\omega)$, especially when the resonator is coupled to its environment primarily by radiation. We therefore assume $H_{{th,eff}}(\omega) \approx H_{th}(\omega)$ in the following.

In an ideal case where all non-fundamental noise sources (i.e., $S_{y,{tmech}} + S_{y,{read}}$) are minimized compared to the fundamental temperature fluctuation noise $S_{y,T}$ (i.e., $S_y \approx S_{y,T}$), NEP reduces to
\begin{equation}
\mathrm{NEP_T} = \frac{\sqrt{4k_B T^2 G}}{\gamma}.
\label{art2_eq6}
\end{equation}
Furthermore, Eq.~\ref{art2_eq6} can be normalized by the sensor area to provide the maximum possible sensor detectivity 
($D^* = A^{1/2} / \mathrm{NEP}$, where $A$ is the sensor area):
\begin{equation}
D_T^* = \gamma \sqrt{\frac{A}{4k_B T^2 G}}.
\label{art2_eq7}
\end{equation}
Another limiting case occurs when the sensor is thermally coupled to its surroundings primarily by radiation on both its faces:
$G \approx 2 \cdot 4 \sigma_{SB} \varepsilon A T^3$,
where $\sigma_{SB}$ is the Stefan–Boltzmann constant, $\varepsilon$ is the total hemispherical emissivity, and a perfect emitter/absorber is assumed ($\varepsilon = \gamma = 1$). In this case, $D^*$ converges to the well-known fundamental limit of thermal photon fluctuation 
($D^*_{T, photon} = 1.3~\mathrm{cm \cdot Hz^{1/2} W^{-1}}$ for a two-side coupled sensor).

Of central importance, we see in Eq.~\ref{art2_eq6}--\ref{art2_eq7} that a thermal sensor limited purely by temperature fluctuation noise $S_{y,T}$ is unaffected by the thermal filter $H_{th}(\omega)$ and can therefore, in principle, have infinitely large sensing bandwidth. More specifically, a sensor can be expected to operate at its fundamental NEP and $D^*$ performance limits (Eq.~\ref{art2_eq6}--\ref{art2_eq7}) at any frequency where its overall noise $S_y(\omega)$ is limited by fundamental temperature fluctuations $S_{y,T}(\omega)$. In this case, the sensor bandwidth is not defined anymore by the thermal corner frequency $\omega_{th} = \tau_{th}^{-1}$, but instead by the limit frequencies ($\omega_{c,read}$ and $\omega_{c,tmech}$) above which other non-fundamental noise sources (respectively $S_{y,read}$ and $S_{y,tmech}$ whose expressions can be found in Supplementary Section S1) begin to dominate over $S_{y,T}$.

\begin{figure}
\centering
\includegraphics[scale=1.07]{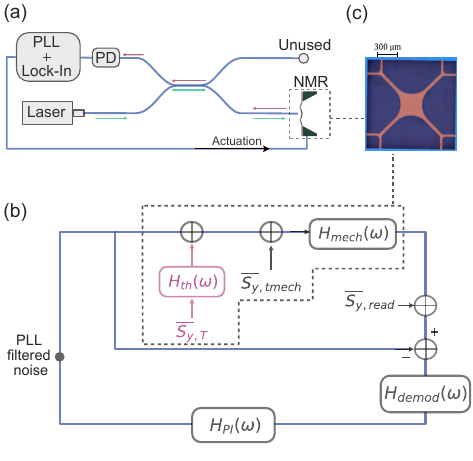}
\caption[Frequency tracking setup schematics]{\textbf{Frequency tracking setup schematics.} 
(a) Schematic of a frequency tracking apparatus for nanomechanical resonators (NMRs), showing a high-level representation of an optical interferometer and a lock-in amplifier equipped with a phase-locked loop (PLL) function. (b) Block diagram illustrating the signal chain for the filtering process of various fractional frequency noise spectral densities (in units of Hz$^{-1/2}$). Here, the noise inputs are unfiltered temperature fluctuation noise $\overline{S_y}_{,T}$, thermomechanical noise $\overline{S_y}_{,tmech}$, and instrument readout noise $\overline{S_y}_{,read}$. The overline donation denotes the white noise magnitude before filtering, i.e., $\overline{S_y}=S_y(0)$. (c) Image of the NMR used in this work.}
\label{art2_fig1}
\end{figure}

To quantify the range of frequencies over which fundamental-level performances are possible, we aim to provide expressions for critical intersection frequencies ($\omega_c$) above which either $S_{y,tmech}$ or $S_{y,read}$ start dominating over $S_{y,T}$. To do so, we must however consider not only the intrinsic noise contributions to $S_y$, but their values after filtering by the frequency tracking apparatus; an optical fiber interferometer combined with a phase-locked loop (PLL) in the present study (see Fig.~\ref{art2_fig1}a). In this setup, an optical fiber tip is positioned in close proximity to the surface of the NMR, creating a low finesse interferometer between the tip of the optical fiber and the NMR. The reflected light signal is collected by a photodetector and subsequently processed by a lock-in amplifier with built-in PLL frequency tracking.

The block diagram in Fig.~\ref{art2_fig1}(b) outlines the noise filtering process when using a PLL to track the resonance frequency. This block diagram configuration is largely inspired from Demir \textit{et al.}~\cite{demir2019fundamental}, with the addition of temperature fluctuation noise $S_{y,T}$. Under the PLL frequency tracking scheme, the filtered thermomechanical noise can be expressed as
\begin{equation}
S^{PLL}_{y,tmech}(\omega) = \frac{2 k_B T}{m_{eff} \omega_r^3 Q A_{rss}^2} \left| H^{PLL}_{mech}(\omega) \right|^2,
\label{art2_eq8}
\end{equation}
where $Q$ is the mechanical quality factor, $\omega_r$ is the resonance frequency at a given eigenmode (in rad/s), $m_{eff}$ is the effective mass, $A_{rss}$ is the resonator driven vibration amplitude (in m), and $H^{PLL}_{mech}(\omega)$ accounts for the PLL filtering effect on thermomechanical noise, as given in~\cite{demir2019fundamental,demir_understanding_2021}:
\begin{equation}
H^{PLL}_{mech}(\omega) = \frac{(j \omega K_p + K_i) H_{demod}(\omega)}{(j \omega)^2 + \frac{j \omega}{\tau_{mech}} + (j \omega K_p + K_i) H_{demod}(\omega)},
\label{art2_eq9}
\end{equation}
where $K_i$ and $K_p$ are the corresponding control loop parameters of the PI controller used in the PLL, $\tau_{mech} = 2Q / \omega_r$ is the mechanical time constant of the NMR, and $H_{demod}(\omega)$ is the lock-in amplifier demodulation filter. The filtered instrument readout noise $S_{read}$ is given by~\cite{schmid2016fundamentals}:
\begin{equation}
S^{PLL}_{y,read}(\omega) = \frac{S_x}{2 Q^2 A_{rss}^2} \left| \frac{H^{PLL}_{mech}(\omega)}{H_{mech}(\omega)} \right|^2,
\label{art2_eq10}
\end{equation}
where $S_x$ is the displacement noise floor of the readout instrument in $\text{m}^2/\text{Hz}$, and $H_{mech}(\omega) = (1 + j \omega \tau_{mech})^{-1}$ is the intrinsic mechanical filter of the resonator.

We model the fundamental temperature fluctuation $S_{y,T}$ as an external noise whereas the intrinsic thermal filter $H_{th}(\omega)$ of the NMR is outside of the signal loop, as in the block diagram in Fig.~\ref{art2_fig1}(b). This is analogous to a step frequency change in the resonator described in~\cite{bevsic2023schemes} and is therefore included in the loop in the same way. In this configuration, the filtered $S_{y,T}$ becomes:
\begin{equation}
S^{PLL}_{y,T}(\omega) = \frac{4 k_B T^2 \alpha^2}{G} \left| H^{PLL}_{mech}(\omega) \cdot H_{th}(\omega) \right|^2.
\label{art2_eq11}
\end{equation}
Conversely, frequency changes resulting from radiation impinging the resonator follow the same signal chain, such that the intrinsic sensor responsivity $R$ is filtered by the PLL, resulting in:
\begin{equation}
R^{PLL}(\omega) = \frac{\gamma \alpha}{G} \left| H^{PLL}_{mech}(\omega) \cdot H_{th}(\omega) \right|.
\label{art2_eq12}
\end{equation}

Note that Eq.~\ref{art2_eq11}--\ref{art2_eq12} corrects our previous model for temperature fluctuation noise in~\cite{zhang2023demonstration}. The conclusions of~\cite{zhang2023demonstration} remain unchanged, but the model proposed herein greatly improves predictability over a broad range of PLL parameters.

\begin{figure}
\centering
\includegraphics[scale=2.2]{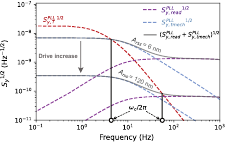}
\caption[Critical intersection frequency]{\textbf{Critical intersection frequency.} Example theoretical frequency fluctuation $S_y$ outlining the critical intersection frequency $\omega_c$ for low and high drive amplitude $A_{rss}$. Resonator parameters are the same as presented in the results section, with a phase-locked loop PLL frequency tracking set at 8 Hz bandwidth, with at demodulation BW set at 5 kHz.  }
\label{art2_fig2}
\end{figure}

We find the limit frequencies ($\omega_{c,read}$ and $\omega_{c,tmech}$) by solving for the intersection between $S^{PLL}_{y,T}$ and, respectively, $S^{PLL}_{y,read}$ or $S^{PLL}_{y,tmech}$, as illustrated in Fig.~\ref{art2_fig2}. The critical frequency above which thermomechanical noise exceeds temperature fluctuation noise is well approximated by:
\begin{equation}
\omega_{c,tmech} \simeq \sqrt{\frac{2 T \alpha^2 m_{eff} \omega_r^3 Q A_{rss}^2}{G}} \cdot \omega_{th},
\label{art2_eq13}
\end{equation}
when the term under the square root is greater than 1. Here, $\omega_{th} = \tau_{th}^{-1}$ is the thermal cut-off frequency of the NMR. Likewise, the frequency above which readout noise exceeds temperature fluctuation noise is approximated as:
\begin{equation}
\omega_{c,read} \simeq \left( \frac{2 k_B T^2 \omega_r^2 \alpha^2 A_{rss}^2}{S_x G} \right)^{1/4} \cdot \sqrt{\omega_{th}}.
\label{art2_eq14}
\end{equation}
Again, assuming the term under the fourth root is greater than 1.

Interestingly, the analytical expressions for $\omega_{c,tmech}$ and $\omega_{c,read}$ indicate that the limit frequencies are independent of the PLL frequency tracking parameters, but rather on parameters intrinsic to the NMR itself (i.e., $\alpha$, $m_{eff}$, $Q$, $A_{rss}$, $\omega_r$, $G$) and the instrument readout noise ($S_x$). In other words, there is no fundamental benefit to adjusting the demodulation bandwidth or PLL control parameter settings to achieve better thermal sensing performances.

Finally, by solving for the intersection between $S^{PLL}_{y,T}$ and $S^{PLL}_{y,read} + S^{PLL}_{y,tmech}$, one can obtain the overall limit frequency $\omega_c$ above which all combined non-fundamental noises exceed the fundamental temperature fluctuation. This is not straightforward analytically, and we instead provide a phenomenological approximation using the analytically obtained $\omega_{c,tmech}$ and $\omega_{c,read}$:
\begin{equation}
\omega_c^{-N} \approx \omega_{c,read}^{-N} + \omega_{c,tmech}^{-N},
\label{art2_eq15}
\end{equation}
which yields good results in the current work for $N \approx 3$, as illustrated in Fig.~\ref{art2_fig2}.

It is worth noting that the overall intersection frequency $\omega_c$ (Eq.~\ref{art2_eq15}) can transition from being dominated by either $\omega_{c,read}$ or $\omega_{c,tmech}$, as we vary the drive amplitude. Even though $S^{PLL}_{y,tmech}$ and $S^{PLL}_{y,read}$ both scale with $A_{rss}^{-2}$ (see Eq.~\ref{art2_eq8} and \ref{art2_eq10}), the intersection equations (Eq.~\ref{art2_eq13}--\ref{art2_eq14}) do not, due to the different $H(\omega)$ filters. As shown in Fig.~\ref{art2_fig2}, for the typical resonator parameters and readout noise used in this work, increasing $A_{rss}$ causes the limit frequency $\omega_c$ to shift from being limited by thermomechanical noise (e.g., for $A_{rss} = 6$ nm) to being limited by readout noise (for $A_{rss} = 120$ nm).

Obviously, driving NMR beyond its mechanical linearity would eventually introduce unwanted frequency fluctuation, originating from Duffing nonlinearity. In this work, we consider the amplitude limit given by Manzaneque \textit{et al.}~\cite{manzaneque2023resolution}:
\begin{equation}
A_{crit} = \frac{2}{\sqrt[4]{3}} \sqrt{\frac{F_s}{\omega_r \beta}},
\label{art2_eq16}
\end{equation}
where $\beta$ is the Duffing coefficient (in m$^{-2}$) and $F_s$ is the sampling frequency of the frequency fluctuations signal. Here, Eq.~\ref{art2_eq16} accounts for the fast-sensing regime in which the intended sampling frequency is much higher than the mechanical cut-off frequency of the NMR (i.e., $\omega > \tau_{mech}^{-1}$). Note that Eq.~\ref{art2_eq16} predicts considerably higher $A_{crit}$ than the more commonly accepted limit \cite{postma2005dynamic} (i.e., $A_{crit} = 0.56 L  (\sigma / QE)^{1/2}$, where $L$ is the size of the NMR, $\sigma$ is the tensile stress and $E$ is the Young’s modulus). For example, setting $F_s \approx 50$ Hz and considering our typical resonator parameters (next section), we obtain $A_{crit} \approx 250$ nm using Eq.~\ref{art2_eq16}.

\section{Methods \& Results}
In this work, we use a low-stress (70 MPa), 90-nm-thick silicon nitride (SiN) NMR (see Fig.~\ref{art2_fig1}c) that has a surface area $A \approx 4.2 \times 10^{-7}~\text{m}^2$. The structure is fabricated via laser ablation~\cite{xie2023laser,nikbakht2023high} of a plain square SiN membrane. Detailed fabrication process and characterization of multiple mechanical eigenmodes are subject of a separate article in preparation. We mount this NMR on a steel plate using three pairs of disk magnets and mechanically excite it with a piezoelectric actuator. All components are placed in a custom-built vacuum chamber operating at pressure $\sim 1 \times 10^{-6}$ Torr to sufficiently suppress air damping, therefore maximizing the NMR $Q$-factor. We measure NMR displacement using a custom-assembled optical fiber interferometer with the configuration shown in Fig.~~\ref{art2_fig1}(a). It consists of a 1564 nm Orion\texttrademark{} distributed feedback (DFB) laser, a 90:10 coupler, a 5 dB optical attenuator, an optical isolator, and a Thorlabs PDA20CS2 photodetector. A Zurich Instrument Ltd. MFLI is used for both exciting and tracking the NMR resonance frequency.

\begin{figure}
\centering
\includegraphics[scale=1.3]{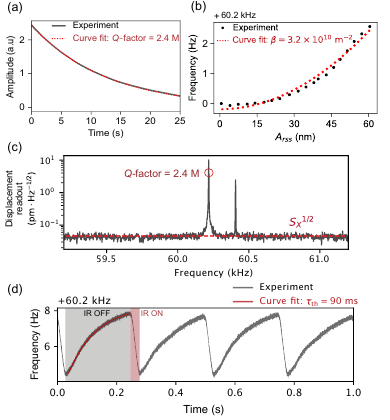}
\caption[General characterization of the resonator mechanical and thermal response]{\textbf{General characterization of the resonator mechanical and thermal response.} (a) Amplitude-time trace of a mechanical ringdown for extracting the NMR mechanical $Q$-factor. (b) Frequency-amplitude plot used for determining the Duffing nonlinearity coefficient of the NMR. (c) Snapshot of the readout instrument displacement noise floor $S_x^{1/2}$, with thermally excited mechanical modes. (d) NMR’s thermal time constant $\tau_{th}$ measurement by intermittently exposing the NMR to broadband IR source modulated using an optical chopper at 4 Hz and 10\% duty cycle.}
\label{art2_fig3}
\end{figure}

We first perform a set of measurements on some of the NMR fundamental thermal and mechanical properties. We pick a low-order mechanical mode at 60.2 kHz that has both a high $Q$-factor ($2.4 \times 10^6$, see Fig.~\ref{art2_fig3}a) and that shows a strong fluctuation response, i.e., that is well aligned with the tip of our optical fiber readout (see Fig.~\ref{art2_fig3}c). Using finite element simulation (see Supplementary Section S3), we find the effective mass $m_{eff}$ for this particular mode is roughly a third of the actual NMR mass $m_0$ (i.e., $m_{eff} = m_0 / 2.9$, where $m_0 = h A \rho$, $h$ is the NMR’s thickness and $\rho$ is SiN material density). Using this mode, we vary the actuation amplitude to measure the Duffing nonlinearity. By fitting the frequency–amplitude data shown in Fig.~\ref{art2_fig3}(b), we extract an experimental Duffing coefficient of $\beta = 3.2 \times 10^{10}~\text{m}^{-2}$.
\begin{figure*}
\centering
\includegraphics[scale=2.5]{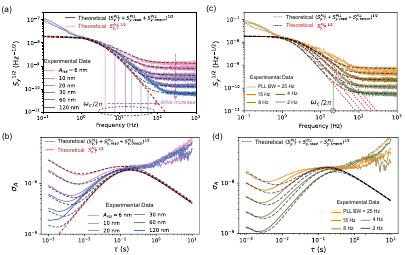}
\caption[Frequency fluctuation measurements]{\textbf{Frequency fluctuation measurements.} (a) Comparison between the experimental fractional frequency spectral density $S_y^{1/2}$ (in colorful solid lines) and their corresponding theoretical predictions (in grey dashed lines), at various vibration amplitude $A_{rss}$ with fixed demodulation bandwidth set to 5 kHz and a phase-locked loop PLL bandwidth of 8 Hz. Theoretical limit frequency $\omega_c/2\pi$ at each $A_{rss}$ is given as a visual guide. (b) Allan deviation $\sigma_A$ at various $A_{rss}$ using the same data as in (a). (c) Comparison between the experimental $S_y^{1/2}$ (in colorful solid lines) and their corresponding theoretical prediction (in grey dashed lines) at varying PLL bandwidth, with fixed $A_{rss}=30$ nm, and a demodulation bandwidth of 5 kHz. (d) Allan deviation $\sigma_A$ at various $A_{rss}$ using the same data as in (c).}
\label{art2_fig4}
\end{figure*}
We also measure the instrument displacement noise floor ($S_x^{1/2} = 0.04~\text{pm} \cdot \text{Hz}^{-1/2}$) as pictured in Fig.~\ref{art2_fig3}(c). This calibration is done in a separate experiment by moving the optical fiber tip towards the membrane by a known distance to calibrate the interferometer sensitivity (in V/m), allowing transformation of raw voltage noise to displacement noise. We measure the NMR thermal response time ($\tau_{th}$) by intermittently exposing the NMR to a broadband IR source modulated using an optical chopper at 4 Hz and 10\% duty cycle. From the fitted transient response (Fig.~\ref{art2_fig3}d), we measure $\tau_{th} = 90$ ms. In this measurement, we set the PLL bandwidth to 40 Hz, which is significantly higher than the thermal cut-off frequency, $\omega_{th}/2\pi = 1.8$ Hz.

The total thermal conductance $G$ between the NMR and its environment, as well as the temperature coefficient of fractional frequency shift $\alpha$ are evaluated by a combination of analytical models and finite-element analysis. We calculate radiative contribution to the thermal conductance using Stefan-Boltzmann law
$G_{rad} = 2 \cdot 4 \sigma_{SB} \varepsilon A T^3 = 4.8 \times 10^{-7}~\text{W/K}$, considering a total emissivity $\varepsilon = 0.1$~\cite{zhang2020radiative}. The factor of 2 accounts for radiative thermal exchange on both front and back NMR surface. Using finite element simulation (see Supplementary Section S2) and considering a thermal conductivity of $2.7~\text{W}~\text{m}^{-1} \text{K}^{-1}$~\cite{snell2022heat} for SiN, we find that $x_{rad} = 65\%$ of the total thermal exchanges with the surrounding environment occurs through radiation (i.e., 35\% via solid-state conduction). The total thermal conductance is then given by $G = G_{rad} / x_{rad} = 7.4 \times 10^7$~W/K. Again, using finite element simulation (see Supplementary Section S3), we obtain a temperature coefficient of fractional frequency shift $\alpha = 6.5 \times 10^{-3}$~K$^{-1}$. Noteworthy, this value is very close (within 1\%) from the value calculated using the analytical expression derived for drum resonators~\cite{zhang2020radiative}:
\begin{equation}
\alpha \simeq \frac{E \alpha_T}{2 \sigma (1 - \nu)},
\label{art2_eq17}
\end{equation}
where $E=300~\mathrm{GPa}$ is the Young's modulus, $\alpha_T=2.2\times10^{-6}~\mathrm{K^{-1}}$ is the thermal expansion coefficient of SiN, $\sigma=70~\mathrm{MPa}$ is the tensile stress and $\nu=0.27$ is the Poisson ratio of SiN. 

When measuring fractional frequency noise of the NMR using PLL, we set the demodulation bandwidth of the lock-in amplifier to a high value (5 kHz), ensuring that noise originating from the readout instrument $S^{PLL}_{y,read}$ remains largely unfiltered. We then set data acquisition rate significantly higher than the demodulation bandwidth (53 kHz) to avoid signal aliasing. We begin by recording frequency-time traces for multiple levels of drive amplitude $A_{rss}$, at a fixed PLL bandwidth of 8 Hz, and thereafter convert those to noise spectral density $S_y^{1/2}$ and Allan deviation $\sigma_A$ (Fig.~\ref{art2_fig4}a--b). The experimental data, in colored lines, closely match our theoretical predictions ($S^{PLL}_{y,tmech} + S^{PLL}_{y,read} + S^{PLL}_{y,T}$) in grey dashed lines, without the need for fitting parameters.

As we increase the drive amplitude ($A_{rss}$), traces increasingly overlap with the predicted $S^{PLL}_{y,T}$ (red dashed line in Fig.~\ref{art2_fig4}a--b), indicating that they are dominated by fundamental temperature fluctuation noise over an increasingly wide bandwidth. To make the measurement bandwidth extension effect visually clearer, in Fig.~\ref{art2_fig4}(a), we include the theoretical overall limit frequency $\omega_c$ for each trace, indicating the point where the noise profile is predicted to transition from the fundamental temperature fluctuation dominated regime to non-fundamental noise dominated regime. In specific, we see that at maximum $A_{rss} = 120$ nm, the NMR can operate at the $S^{PLL}_{y,T}$ dominated regime up to slightly above 54 Hz---roughly two orders of magnitude greater than the NMR’s intrinsic thermal cutoff frequency $\omega_{th}/2\pi = 1.8$ Hz. In contrast, this measurement bandwidth enhancement is not seen in Fig.~\ref{art2_fig4}(c--d) where we solely vary the PLL bandwidth while using a fixed $A_{rss}$ of 30 nm. This aligns with our theoretical prediction, as Eq.~\ref{art2_eq13}--\ref{art2_eq14} are independent of the PLL bandwidth.

In the noise traces of Fig.~\ref{art2_fig4}, we observe some discrepancy between the experiment and the theoretical prediction. At low sampling frequency $\omega \leq \omega_{th}$, we observe the typically unavoidable additional noise due to systematic drift. At higher sampling frequencies ($\omega > \omega_{th}$), we also observe some discrepancy between the theoretical and experimental roll-off in $S^{PLL}_{y,T}$. This is especially visible for higher $A_{rss}$ traces. We hypothesize that this discrepancy arises from our assumption of $H_{th,eff}(\omega) \approx H_{th}(\omega)$ in our theoretical calculation. As hinted in~\cite{kanellopulos2025comparative}, averaging local energy fluctuations into an effective $H_{th,eff}(\omega)$ thermal filter may result in a greater noise bandwidth than the thermal response measured under uniform heating (i.e., $H_{th,eff}(\omega) > H_{th}(\omega)$). This phenomenon warrants quantitative experimental investigation in future work.

Readout laser back-action~\cite{sadeghi2020frequency,kanellopulos2025comparative} and Duffing nonlinearity~\cite{manzaneque2023resolution} are also known to cause unwanted frequency fluctuations, but those are ruled out in the present case. We verify that laser power fluctuation plays a negligible role in this experiment by repeating measurements with a variable optical attenuator (see Supplementary Section S4). We also observe that the frequency fluctuation originating from Duffing nonlinearity is negligible. We note in Fig.~\ref{art2_fig4}(a) that higher $A_{rss}$ never increases noise. High $A_{rss}$ either reduces noise at very high frequencies (where additive noise dominate) or has no effect on the noise (at frequencies where temperature fluctuations noise $S_{y,T}$ dominate). Negligible nonlinear noise is consistent with the critical amplitude predicted by Eq.~\ref{art2_eq16}. We are mostly interested in $\omega/2\pi$ with sampling frequencies $F_s$ between 4 to 54 Hz. By using our measured $\beta$ value into Eq.~\ref{art2_eq16}, we estimate $A_{crit}$ increases from 70 to 255 nm for this range of $F_s$, which is generally above our experimental $A_{rss}$.

\begin{figure}
\centering
\includegraphics[scale=0.65]{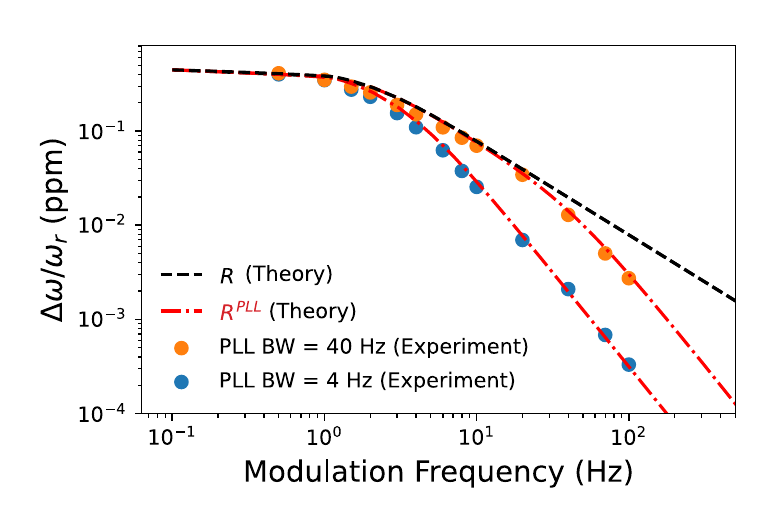}
\caption[Thermal responsivity measurement]{\textbf{Thermal responsivity measurement.} Circular dots represent experimental fractional frequency shift $\Delta\omega/\omega_r$ induced by impinging thermal radiation at various modulation frequencies under closed-loop frequency tracking. Blue dots represent data with phase-locked loop PLL set to 4 Hz and a demodulation bandwidth of 5 kHz. Orange dots represent with PLL set to 40 Hz and a demodulation bandwidth of 5 kHz. Red dashed lines represent theoretical prediction (Eq.~\ref{art2_eq12}) corresponding to each experimental trace. Black dashed line represents the modulated theoretical thermal responsivity curve without closed-loop frequency tracking (i.e., without PLL).}
\label{art2_fig5}
\end{figure}

Another important aspect to confirm the validity of our model is to confirm the predicted effect of PLL filtering on the NMR’s thermal responsivity, i.e., $R^{PLL}$. In Fig.~\ref{art2_fig5}, we record traces of the NMR’s responsivity at largely different PLL bandwidths (4 and 40 Hz). For each trace, we modulate the thermal source from 0.5 Hz to 100 Hz. The experimental data closely matches theoretical prediction from Eq.~\ref{art2_eq12} (red dashed lines) in both cases. In contrast, the black dashed line in Fig.~\ref{art2_fig5} includes only the NMR’s intrinsic thermal filter $H_{th} = (1 + j \omega \tau_{th})^{-1}$ which, as expected, does not align with our measurement.
\begin{figure}
\centering
\includegraphics[scale=2.5]{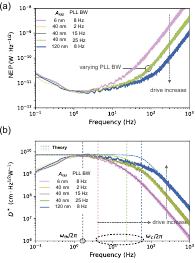}
\caption[Enhanced NEP and $D^*$ bandwidth]{\textbf{Enhanced NEP and $D^*$ bandwidth.} Experimental noise equivalent power NEP (a) and detectivity (b) of the nanomechanical resonator at various vibration amplitudes $A_{rss}$ and phase-locked loop PLL bandwidths. As predicted, the critical frequency ($\omega_c$) at which sensor performance decline significantly is greater than the thermal cutoff frequency ($\omega_{th}$), by up to a factor 30. The curves also confirm that $\omega_c$ depends strongly on the drive amplitude ($A_{rss}$), but not on the frequency tracking PLL bandwidth (the curves at different PLL completely overlap and appear green). Experimental specific detectivity $D^*$ traces are obtained from the NEP values in (a) using $\sqrt{A}/\text{NEP}$, where $\sqrt{A} = 300~\mathrm{\mu m}$ is the characteristic size of the center pad of the NMR. Theoretical predictions for $D^*$ are shown in dashed lines with matching colors. These predictions assume $\gamma = 0.4$ as the absorption, which corresponds to the material absorption of 100 nm thick SiN at $\lambda = 11~\mu\text{m}$ wavelength.}
\label{art2_fig6}
\end{figure}

Lastly, we obtain our sensor NEP and $D^*$ by combining our experimentally measured noise $S_y$ (i.e., Fig.~\ref{art2_fig4}) with our experimentally validated responsivity model (Fig.~\ref{art2_fig5}). We assume $\gamma = 0.4$, which corresponds to the peak material absorption of 90-nm-thick SiN membrane at a wavelength $\lambda = 11~\mu\text{m}$~\cite{zhang2020radiative}. Of course, in a practical setting, this material absorption peak would be too narrow, and a broadband optical absorber would be required. We expect this step to be straightforward since previous work by our group and others~\cite{zhang2024highdetectivityterahertzradiation, piller2022thermal} have demonstrated that including such radiation absorber does not degrade the sensor mechanical performances.

We present both NEP and $D^*$ in Fig.~\ref{art2_fig6} for various $A_{rss}$ and PLL bandwidth. In all cases, our calculated $\omega_c$ values (using Eq.~\ref{art2_eq15}) accurately predict the corner frequency at which experimental $D^*$ declines. For the highest $A_{rss}$ drive, this decline occurs at 54 Hz, i.e., at a 30 times higher frequency than the sensor intrinsic thermal bandwidth ($\omega_{th}/2\pi = 1.8$ Hz). As a visual guide, in Fig.~\ref{art2_fig6}(b), we also include both theoretical $D^*$ and theoretical overall limit frequency $\omega_c$ for each $A_{rss}$. We observe strong agreement between the theoretical $D^*$ traces. The minor discrepancies are due to the same reasons discussed in the context of Fig.~\ref{art2_fig4}.

As expected also from our theoretical model, PLL bandwidth have negligible influence on the NEP and $D^*$. At a fixed $A_{rss} = 40$ nm drive, we observe that varying the PLL bandwidth from 2 to 25 Hz leads to no difference in sensing performance as the three traces are perfectly superimposed. PLL filtering affects both the responsivity and the noise in the same manner and therefore cancels out in the final NEP and $D^*$.

\section{Conclusion}

In conclusion, we have demonstrated an NMR radiation sensor limited by temperature fluctuation noise over an unprecedented measurement bandwidth (i.e., $30 \times \omega_{th}$). This demonstration opens avenues for developing fast NMR-based thermal sensors operating at the fundamental detectivity $D^*$ limit at room temperature, which could serve as next-generation long-wavelength infrared (LWIR) thermal sensors. An obvious next step for this work is to demonstrate the same signal bandwidth enhancement in a sensor comprising a broadband optical absorber, rather than relying solely on the intrinsic material absorption of SiN. This should be straightforward since previous work has demonstrated that including such a broadband radiation absorber does not degrade the sensor mechanical performances.
\newpage
\nocite{*}
\bibliography{apssamp}

\newpage
\vspace{1cm}
\section*{S1. Expressions for intrinsic thermomechanical and instrument readout noise}
Here, we provide expressions for intrinsic thermomechanical noise ($S_{y,\mathrm{tmech}}$) and instrument readout noise ($S_{y,\mathrm{read}}$) (i.e., without the loop dynamics shown in Fig.~\ref{art2_fig1}). The thermomechanical noise in the open-loop case can be expressed as:
\begin{equation*}\label{fun_eq:15} S_{y,tmech}(\omega)=\frac{2k_BT}{m_{eff}\omega_r^3QA_{rss}^2}|H_{mech}(\omega)|^2,  \tag{S1}
\end{equation*}
where $H_{mech}(\omega)$ is the intrinsic mechanical filter of the NMR, calculated as $1/(1+j\omega\tau_{mech})$ where $\tau_{mech}$ is the mechanical time constant. The instrument readout noise can be expressed as: 
\begin{equation*}\label{fun_eq:16} S_{y,read}(\omega)=\frac{S_x}{2Q^2A_{rss}^2}|H_{demod}(\omega)|^2,  \tag{S2}
\end{equation*}
where $H_{demod}(\omega)$ accounts for the readout instrument filter (e.g., lock-in amplifier demodulation filter), calculated as $1/(1+j\omega\tau_{demod})$ where $\tau_{demod}$ represents the demodulation speed.

\section*{S2. Heat transfer simulation}
\begin{figure}[h!]
\centering
\includegraphics[scale=0.26]{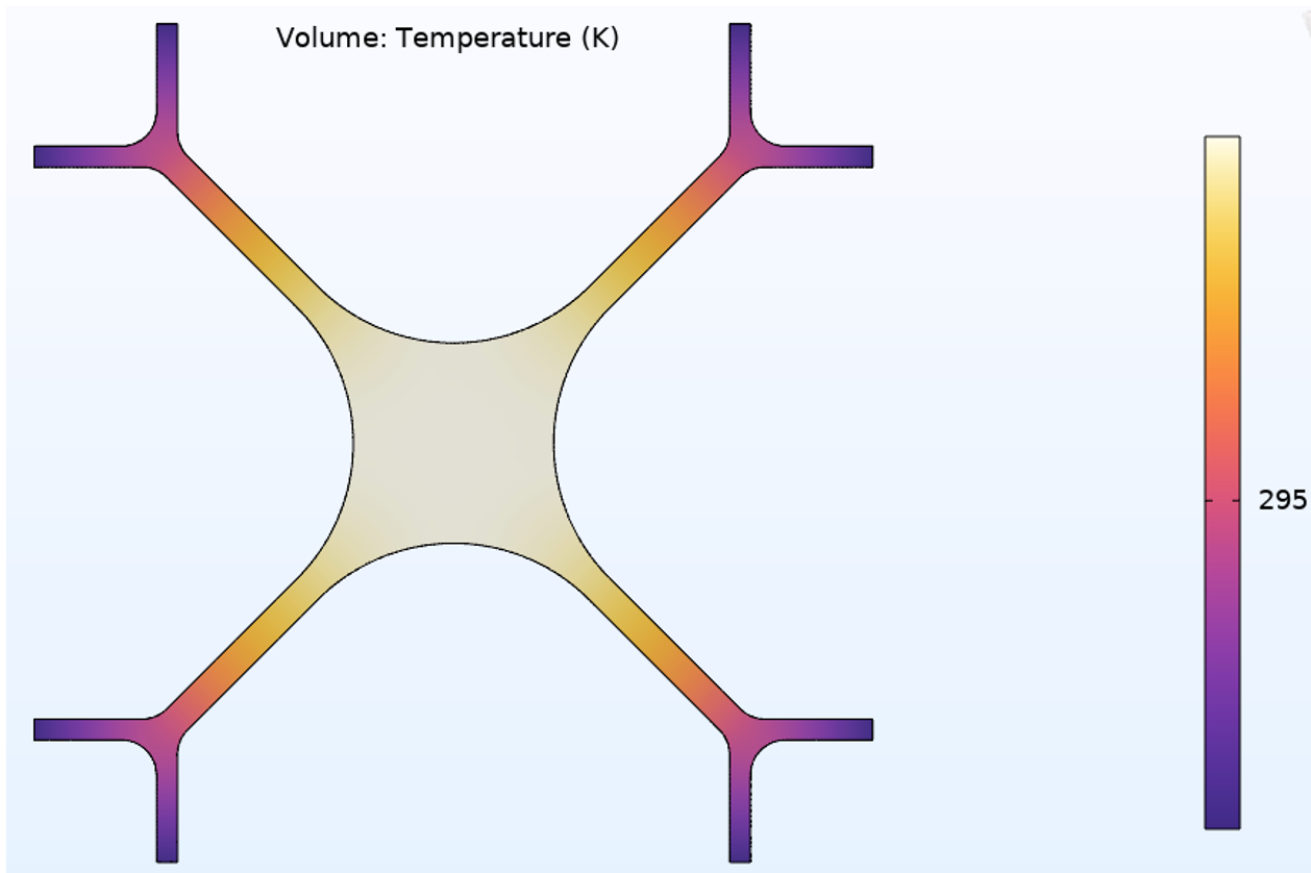}

\vspace{0.5em}
\justifying
{\textbf{Fig.~S1.} 2-D temperature profile of the SiN nanomechanical resonator
considering solid-state conduction and radiative heat transfer during
1~$\mu$W constant heating.\par}
\end{figure}
We perform a heat transfer simulation via COMSOL, considering both solid-state conduction and radiative heat transfer. We apply an uniform heat flux of $Q = 1~\mu\text{W}$ being absorbed by the SiN nanomechanical resonator (NMR) surface and set the boundary condition at constant room temperature $T_{room} = 294.15$~K. Fig.~S1 shows the 2-D temperature profile in steady state during constant heating.

In COMSOL, we obtain heat flux leaving the NMR via solid-state conduction for a single tether to be 
$Q_{cond,1~tether} = 43.425~\text{nW}.$ The total heat flux leaving the NMR via conduction can be calculated as $Q_{cond} = 8 \times 43.425 = 347.4~\text{nW}$ in which the factor of 8 accounts for the total number of tethers supporting the NMR. Due to only two heat transfer modes (e.g., conduction and radiation) included in this simulation, we can calculate the percentage of heat flux leaving the NMR via radiation as:
\begin{equation*}
x_{rad} = \frac{Q - Q_{cond}}{Q} = \frac{1 \times 10^{-6} - 347.4 \times 10^{-9}}{1 \times 10^{-6}} \approx 65\%. \tag{S3}
\end{equation*}

\section*{S3. Effective mass and temperature coefficient of fractional frequency shift simulation}

\begin{figure}[h!]
\centering
\includegraphics[scale=0.22]{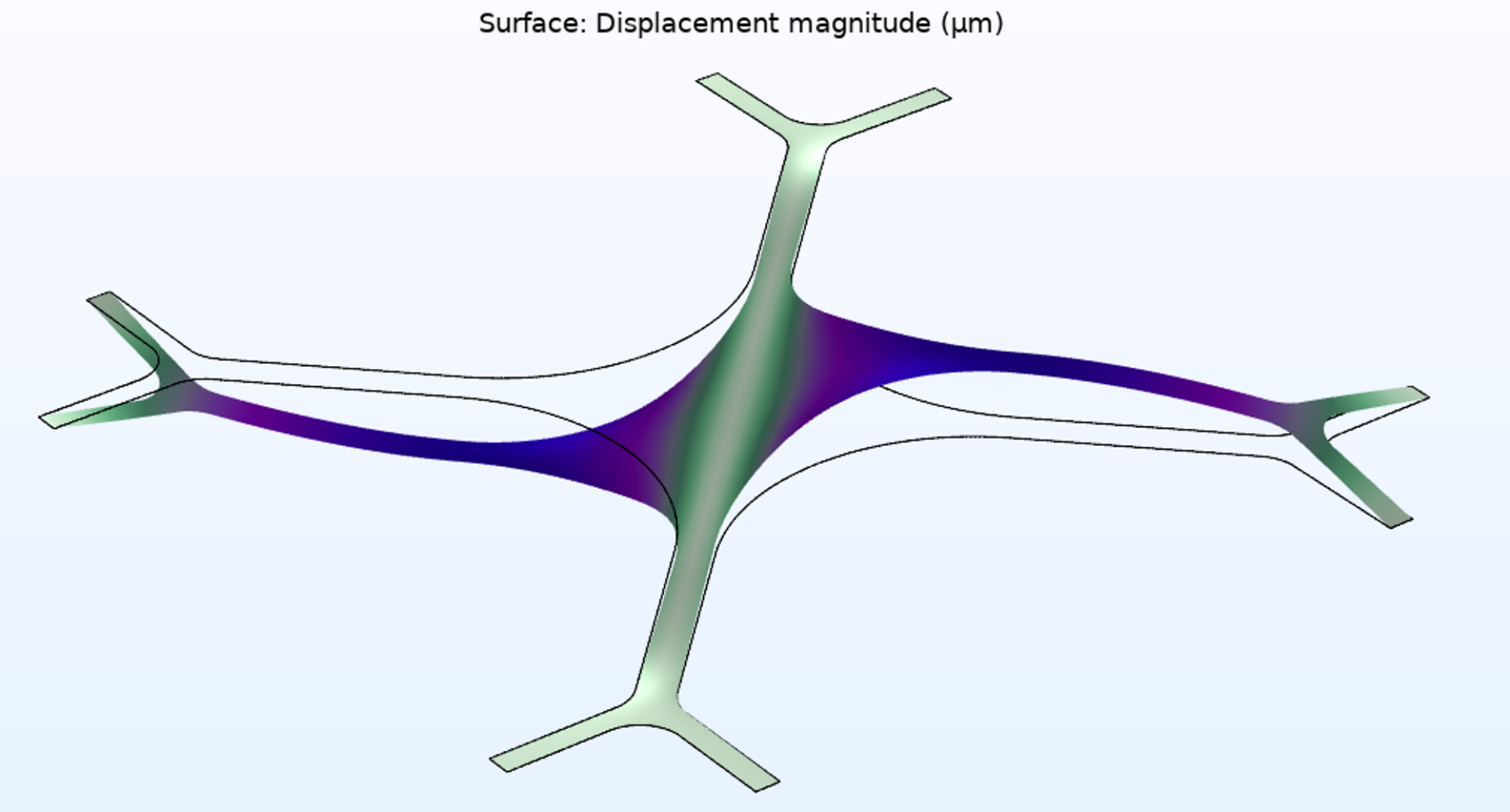}

\vspace{0.5em}
\justifying
{\textbf{Fig.~S2.} Mechanical mode (1,2) of the SiN nanomechanical resonator in COMSOL simulation.\par}
\end{figure}

Using SiN material density $\rho = 2900~\text{kg/m}^3$, and the dimension of the fabricated NMR, we perform dynamic vibration analysis in COMSOL which solves for eigen modes and their corresponding eigen frequencies $f_r$. According to this simulation, at room temperature $T_0 = 294.15~\text{K}$, $f_r$ of mechanical mode (1,2) is 60.8 kHz (see Fig.~S2) when the tensile stress of the SiN membrane is set to $\sigma = 70~\text{MPa}$. This closely aligns with the chosen mode in our experiment ($f_r = 60.2~\text{kHz}$, see main text). From this simulation, we also obtain an effective mass, $m_{eff} \approx 36.5$ ng for mode (1,2). This value is roughly a third of the actual mass, $m_0 = h A \cdot \rho$ of the NMR (i.e., $m_{eff} = m_0 / 2.9$).

By applying uniform heating on the surface of the NMR in COMSOL until its average temperature increases by $\approx 1$~K, we observe $f_r$ of the mechanical mode (1,2) decreases by 393 Hz. From this, we extract the temperature coefficient of fractional frequency $\alpha \approx 6.45 \times 10^{-3}~\text{K}^{-1}$. Intriguingly, this value matches closely (i.e., $<1$\% discrepancy) with the commonly used equation for estimating $\alpha$ of the fundamental mechanical mode of a drum resonator under tensile stress, listed as follows:
\begin{equation*}
\alpha = \frac{E \alpha_T}{2 \sigma (1 - \nu)}. \tag{S4}
\end{equation*}
We use Young’s modulus $E = 300$~GPa, Poisson’s ratio: $\nu = 0.27$, thermal expansion coefficient $\alpha_T = 2.2 \times 10^{-6}~\text{K}^{-1}$ and tensile stress $\sigma = 70$~MPa for this work.

\section*{S4. Frequency fluctuation under varying readout laser power}

The power intensity fluctuation of the probing laser is known to induce unwanted frequency noise to the temperature-sensitive NMR. This effect arises from the probing laser acting as a continuous, point heat source on the NMR. When the laser power fluctuates, it induces frequency fluctuation whose magnitude scales with both the laser power and the NMR’s thermal responsivity $R$ (see Eq.~\ref{art2_eq2} in main text), on the timescale set by NMR’s thermal time constant $\tau_{th}$.

To investigate this, we use a variable optical attenuator (Thorlabs V1550A) to vary the incident laser power on the NMR for multiple traces of frequency noise measurement. In Fig.~S3, we present frequency noise measurements at different laser power attenuation levels, ranging from 0.25 to 10 dB. We see no increase of the NMR noise when reducing attenuation below the value used in the main text (5 dB). In turn, a low attenuation (i.e., high laser power) helps improve the displacement sensitivity $S_x$ of the readout instrument, resulting in a reduction in readout noise $S^{PLL}_{y,read}$ at higher sampling frequency ($f \gg 1 / 2 \pi \tau_{th}$).

\begin{figure}[h!]
\centering
\includegraphics[scale=0.33]{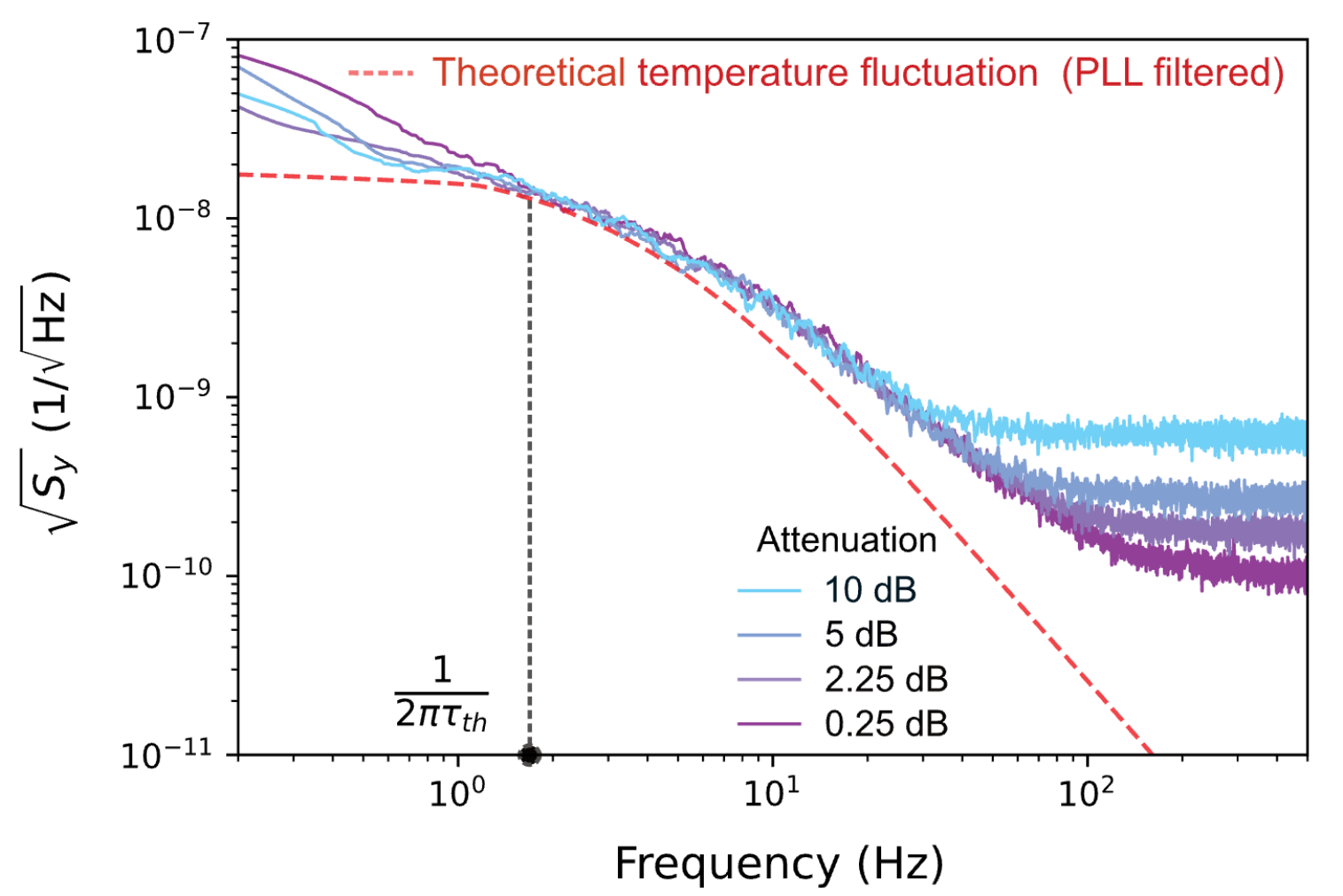}

\vspace{0.5em}

{\justifying \textbf{Fig.~S3.} Frequency fluctuation measurement at fixed vibration amplitude $A_{rss}$ and phase-locked loop settings (i.e., PLL and demodulation bandwidth) with varying laser power. 5dB attenuation is used in the main text.\par}
\end{figure}

\end{document}